\documentclass[10pt,A4paper,amsmath,amssymb,aps,etoolbox,floatfix,longbibliography,nofootinbib,
showkeys,noshowpacs,superscriptaddress,twocolumn]{revtex4}

\newcommand{\bea}{\begin{eqnarray}}
\newcommand{\beq}{\begin{equation}}

\newcommand{\eea}{\end{eqnarray}}
\newcommand{\eeq}{\end{equation}}
\newcommand{\ds}{\displaystyle}

\usepackage{amsmath}
\usepackage{amsfonts}
\usepackage{amssymb}
\usepackage{bm}       
\usepackage[usenames,dvipsnames]{color}
\usepackage{comment}
\usepackage{dcolumn}  
\usepackage{epsfig}
\usepackage{graphics}
\usepackage{graphicx} 
\usepackage{lipsum}
\usepackage{natbib}
\usepackage[rightcaption]{sidecap}

\usepackage{placeins} 
\usepackage{times}
\usepackage[varg]{txfonts}
\usepackage{xcolor}
\usepackage{xspace}

\graphicspath{%
    {converted_graphics/}
    {/}
}

\begin{document}

\title{Testing the Amp\`ere-Maxwell law on the photon mass and Lorentz symmetry violation with MMS multi-spacecraft data}

\author{Alessandro D.A.M. Spallicci \footnote{email: spallicci@cnrs-orleans.fr 
}}

\affiliation{\mbox{Laboratoire de Physique et Chimie de l'Environnement et de l'Espace (LPC2E) UMR 7328}\\
\mbox{Centre National de la Recherche Scientifique (CNRS), Universit\'e d'Orl\'eans (UO), Centre National d'\'Etudes Spatiales (CNES)}\\
\mbox {3A Avenue de la Recherche Scientifique, 45071 Orl\'eans, France}}

\affiliation{\mbox{Institut Denis Poisson (IDP) UMR 7013}\\
\mbox{Universit\'e d'Orl\'eans (UO) et Universit\'e de Tours (UT), Centre National de la Recherche Scientifique (CNRS)}\\
\mbox{Parc de Grandmont, 37200 Tours, France}}

\affiliation{\mbox{UFR Sciences et Techniques, 
Universit\'e d'Orl\'eans, Rue de Chartres, 45100 Orl\'{e}ans, France}}

\affiliation{\mbox{Observatoire des Sciences de l'Univers en region Centre (OSUC) UMS 3116} \\
\mbox{ Universit\'e d'Orl\'eans (UO), Centre National de la Recherche Scientifique (CNRS)}
\mbox{ Observatoire de Paris (OP), Universit\'e Paris Sciences \& Lettres (PSL)}\\
\mbox{1A rue de la F\'{e}rollerie, 45071 Orl\'{e}ans, France}}

\author{Giuseppe Sarracino \footnote{email: giuseppe.sarracino@inaf.it}}

\affiliation{\mbox {Osservatorio Astronomico di Capodimonte (OACN), Istituto Nazionale di Astrofisica (INAF)}\\
\mbox {Salita Moiariello 16, 80131 Napoli, Italy}}

\affiliation{\mbox {Dipartimento di Fisica E. Pancini, Universit\`a degli Studi di Napoli Federico II (UNINA), and}\\ 
\mbox {Istituto Nazionale di Fisica Nucleare (INFN), Sezione di Napoli}\\
\mbox {Complesso Universitario Monte S. Angelo, Via Cinthia 9 Edificio G, 80126 Napoli, Italy}}

\author{Or\'elien Randriamboarison \footnote{email: orelien.randriamboarison@cnrs-orleans.fr}} 

\affiliation{\mbox{Laboratoire de Physique et Chimie de l'Environnement et de l'Espace (LPC2E) UMR 7328}\\
\mbox{Centre National de la Recherche Scientifique (CNRS), Universit\'e d'Orl\'eans (UO), Centre National d'\'Etudes Spatiales (CNES)}\\
\mbox {3A Avenue de la Recherche Scientifique, 45071 Orl\'eans, France}}

\affiliation{\mbox{UFR Sciences et Techniques, 
Universit\'e d'Orl\'eans, Rue de Chartres, 45100 Orl\'{e}ans, France}}

\affiliation{\mbox{Observatoire des Sciences de l'Univers en region Centre (OSUC) UMS 3116} \\
\mbox{ Universit\'e d'Orl\'eans (UO), Centre National de la Recherche Scientifique (CNRS)}
\mbox{ Observatoire de Paris (OP), Universit\'e Paris Sciences \& Lettres (PSL)}\\
\mbox{1A rue de la F\'{e}rollerie, 45071 Orl\'{e}ans, France}}

\author{Jos\'e A. Helay\"el-Neto \footnote{email: helayel@cbpf.br}}

\affiliation{\mbox{Departamento de Astrof\'{\i}sica, Cosmologia e Intera\c{c}\~{o}es Fundamentais (COSMO), Centro Brasileiro de Pesquisas F\'{\i}sicas (CBPF)}\\
\mbox {Rua Xavier Sigaud 150, 22290-180 Urca, Rio de Janeiro, RJ,  Brazil}}

\author{Abedennour Dib \footnote{email: abedennour.dib@cnrs-orleans.fr }}

\affiliation{\mbox{Laboratoire de Physique et Chimie de l'Environnement et de l'Espace (LPC2E) UMR 7328}\\
\mbox{Centre National de la Recherche Scientifique (CNRS), Universit\'e d'Orl\'eans (UO), Centre National d'\'Etudes Spatiales (CNES)}\\
\mbox {3A Avenue de la Recherche Scientifique, 45071 Orl\'eans, France}}

\affiliation{\mbox{UFR Sciences et Techniques, 
Universit\'e d'Orl\'eans, Rue de Chartres, 45100 Orl\'{e}ans, France}}

\affiliation{\mbox{Observatoire des Sciences de l'Univers en region Centre (OSUC) UMS 3116} \\
\mbox{ Universit\'e d'Orl\'eans (UO), Centre National de la Recherche Scientifique (CNRS)}
\mbox{ Observatoire de Paris (OP), Universit\'e Paris Sciences \& Lettres (PSL)}\\
\mbox{1A rue de la F\'{e}rollerie, 45071 Orl\'{e}ans, France}}

\begin{abstract}
We investigate possible evidence from Extended Theories of Electro-Magnetism by looking for deviations from the Amp\`ere-Maxwell law. The photon, main messenger for interpreting the universe, is the only free massless particle in the Standard-Model (SM). Indeed, the deviations may be due to a photon mass for the de Broglie-Proca (dBP) theory or the Lorentz Symmetry Violation (LSV) in the SM Extension (SME), but also to non-linearities from theories as of Born-Infeld, Heisenberg-Euler. With this aim, we have analysed six years of data of the Magnetospheric Multi-Scale mission, which is a four-satellite constellation, crossing mostly turbulent regions of magnetic reconnection and collecting about 95\% of the downloaded data, outside the solar wind. We examined 3.8 million data points from the solar wind, magnetosheath, and magnetosphere regions. In a minority of cases, for the highest time resolution burst data and optimal tetrahedron configurations drawn by the four spacecraft, deviations have been found ($2.2\%$ in modulus and $4.8\%$ in Cartesian components for all regions, but raising up in the solar wind alone to $20.8\%$ in modulus and $29.7\%$ in Cartesian components and up to 45.2\% in the extreme low-mass range). The deviations might be due to unaccounted experimental errors or, less likely, to non-Maxwellian contributions, 
for which we have inferred the related parameters for the dBP and SME cases. Possibly, we are at the boundaries of measurability for non-dedicated missions. We discuss our experimental results (upper limit of photon mass of $2.1 \times 10^{-51}$ kg, and of the LSV parameter $|\vec{k}^{\rm AF}|$ of $6 \times 10^{-9}$ m$^{-1}$), as the deviations in the solar wind, versus more stringent but model-dependent limits. 

\end{abstract}

\date{12 March 2024}


\maketitle


\section{Beyond Maxwellian theory: implications}

Astrophysical observations are mostly based on electro-magnetic signals, interpreted with the standard Maxwellian massless and linear framework. Even when considering its overwhelming successes, the XIX century Maxwellian theory might be an approximation of a more comprehensive theory, as Newtonian gravity is for General Relativity. The linear description of the electro-magnetic phenomena by Maxwell's theory and the unique photon masslessness, among all particles in the Standard-Model (SM) are questioned by the Extended Theories of Electro-Magnetism (ETEM). Herein, we test the Amp\`ere-Maxwell (AM) law, by analysing millions of measured currents in the Earth neighbourhood. 

The search for a deviation from the AM law complements other searches of manifestations or implications of ETEM physics among which

\begin{itemize}
\item{vacuum birefringence: the refractive index is dependent on the polarisation and propagation direction  \cite{kostelecky-mewes-2009,Li-etal-2015};}
\item {frequency shifts associated with magnetars from non-linear quantum electro-dynamics  \cite{MosqueraCuesta-Salim-2004b,bopbsp2017};}
\item {light dispersion: the group velocity differs from $c$ by a factor proportional to the inverse square of the frequency \cite{boelmasasgsp2016,boelmasasgsp2017} with a bearing on i) multi-messenger astronomy \cite{Veske-etal-2021}, ii)on pulsar timing \cite{Donner-etal-2019} and on gravitational wave detection at low frequencies via pulsars \cite{Antoniadis-etal-2023}, due to plasma dispersion \cite{bebosp2017}, iii) and on the estimates of the graviton mass \cite{Piorkowska-Kurpas-2022};}
%
%
\item {frequency shift {\it in vacuo}  in addition  to the cosmological redshift due to the expansion of the Universe \cite{helayelneto-spallicci-2019,spallicci-etal-2021,spallicci-sarracino-capozziello-2022,sarracino-spallicci-capozziello-2022}; 
thanks to this addition, a concordance between the luminosity 
and the redshift distances of Supernovae IA has been
achieved, without requiring an accelerated expansion due to dark energy; for a complementary approach, see \cite{LopezCorredoira-CalvoTorel-2022};}   
\item {massive photons were related to dark energy \cite{Kouwn-Oh-Park-2016}, whereas dark photons, massive and massless, were related to dark matter \cite{Landim-2020,fabbrichesi-gabrielli-lanfranchi-2021, caputo-etal-2021b}; dark photons are differently connected to the SM; they possess a small kinetic mixing with the visible photon at low energies leading to oscillations; the damping of dark (DP) and massive photons (MP) are respectively $-{\ds }(1 + \chi e^{\ds {r}/{\lambda_{\rm DP}}})/r$ and $-{\ds}(e^{\ds {r}/{\lambda_{\rm MP}}})/r$; this difference is not necessarily detectable if $\chi$, representing the mixing, and the masses, are both small; in \cite{caputo-etal-2021b}, bounds on dark photons have been obtained via constraints on the photon mass; their detectability through black holes is discussed in \cite{caputo-etal-2021a};}
\item {massive photon gasses considered for cosmology \cite{cuzinatto-etal-2017}.}
\end{itemize}


\section{Extended Theories of Electro-magnetism (ETEM)}

De Broglie \cite{debroglie-1923} first proposed a massive photon, estimated below $10^{-53}$ kg through dispersion analysis. He supervised the doctorate work of A. Proca who wrote \cite{proca-1936d,proca-1937} a generic Lagrangian for electrons, positrons, neutrinos, and photons, asserting the masslessness of the latter two, being the photon composed of two pure charge massless particles, see Appendix (\ref{proca}). Instead, de Broglie expected the photon being composed of two oppositely charged massive particles and he was the first to write  the Maxwell massive equations \cite{debroglie-1936,db40}. These were obtained through a complex theory of mutually corresponding Dirac particles and electro-magnetic operators, and not through the principle of minimal action; see also \cite{debroglie-1950}. The de Broglie-Proca (dBP) theory is not gauge-invariant (a change in the potential implies a change in the field), but it satisfies Lorentz Symmetry (LoSy) according to which measurements are independent of the orientation or the velocity of the observer. Later other formulations of massive theories showed gauge-invariance \cite{bopp-1940,podolski-1942,stueckelberg-1957}. For renormalisation, unitarity, origin of mass, and charge conservation, see \cite{boulware-1970,guendelman-1979,nussinov-1987,itzykson-zuber-2006,addvgr-2007,scharffgoldhaber-nieto-2010}; 
for recent discussions on modified electro-dynamics, see for 
various aspects of LoSy violation (LSV) \cite{
casana-ferreira-lisboasantos-dossantos-schreck-2018,
felipe-etal-2019,
Reis-etal-2019,
Ferreira-etal-2020,
arajuio-maluf-2021,
silva-etal-2021,
marques-etal-2022,
terin-spalenza-belich-helayelneto-2022}, 
for higher derivative terms \cite{Ferreira-etal-2019},
for massive photons in the swampland \cite{craig-garciagarcia-2018,reece-2019},
for very special relativity \cite{alfaro-soto-2019}, 
for massive Quantum-Electrodynamics (QED) \cite{govindarajan-kalyanapuram-2019},
for axion-photon mixing \cite{paixao-ospedal-neves-helayelneto-2022}.

The successful SM is LoSy conform, but it does not explain neutrino oscillations and masses, the matter-antimatter unbalance, nor it provides particle candidates for dark matter \cite{Ellis-2009}, nor it includes a new interaction for dark energy, and possibly faces the W Boson mass issue \cite{Aaltonen-etal-2022}. 

Thus, theories beyond the SM were proposed: the SM Extension (SME) \cite{colladay-kostelecky-1997,colladay-kostelecky-1998}, based on 
LSV \cite{Tasson-2014}.
The SME leads to a gauge-invariant, but anisotropic photon mass \cite{bonetti-dossantosfilho-helayelneto-spallicci-2017,bonetti-dossantosfilho-helayelneto-spallicci-2018} proportional to the LSV parameters, represented by a $k^{\rm AF}_{\alpha}$ 
4-vector for the odd handedness of the 
Charge conjugation-Parity-Time reversal (CPT) symmetry and by a $k_{\rm F}^{\alpha\nu\rho\sigma}$ 
tensor for even CPT. 

$k^{\rm AF}_{\alpha}$ from the Carroll-Field-Jackiw (CFJ) Lagrangian \cite{carroll-field-jackiw-1990} induces a mass, while $k_{\rm F}^{\alpha\nu\rho\sigma}$ only in Super-Symmetry after photino integration \cite{bonetti-dossantosfilho-helayelneto-spallicci-2017,bonetti-dossantosfilho-helayelneto-spallicci-2018}. 
This confirms previous work \cite{bsbebohn2003,baetaetal2004} that showed that LoSy breaking modifies the propagators to include a massive mode proportional to the violation vector, see Appendix (\ref{LSV-SME}).

Belonging to the ETEM framework, Non-Linear Electro-Magnetism (NLEM) was proposed first by Born and Infeld (BI) 
\cite{born-infeld-1934a,born-infeld-1935} for regularising point charges and by Heisenberg and Euler (HE) \cite{heisenberg-euler-1936} for dealing with strong fields. Nowadays, both theories are used for second order QED. For a review including more recent theories, see \cite{Sorokin-2022}. Whether in the NLEM framework the photon acquires a mass is yet unanswered.

The existence of an electro-magnetic damping in {\it vacuo} hints towards the idea that the electro-magnetic interaction does not have an infinite range anymore. This is an expected behaviour of massive models.
A Boson of mass $m$ damps the interaction at a distance $r$ by a factor $e^{- r/\lambdabar}$, $ \lambdabar~{\rm [m]} = 
\hbar/cm = 3.52 \times 10^{-43}/ m~{\rm [kg]}$ being the reduced Compton length. 

The lowest measurable value for any mass is dictated by Heisenberg's principle $m \geq \hbar/\Delta t c^2$, and gives $1.3\times 10^{-69}$ kg,
where $\Delta t$ is the supposed age of the Universe \cite{capozziello-benetti-spallicci-2020, spallicci-benetti-capozziello-2022}. The corresponding reduced Compton length is $2.6 \times 10^{26}$ m. 

Thus, tinier photon masses may be sought through large-scale measurements \cite{scharffgoldhaber-nieto-2010}. We will analyse the AM law, through the NASA Magnetospheric Multiscale Mission (MMS) \cite{Fuselier-etal-2016}, with a method previously applied to the ESA Cluster mission \cite{retino-spallicci-vaivads-2016}. MMS was launched on 13 March 2015, and it is composed of four identical satellites flying in a tetrahedral formation, passing through different Earth neighbourhoods, carrying particle detectors and magnetometers. 

We use SI units ($1$ kg $ = 5.61 \times 10^{35} $ eV in natural units) and the Geocentric Solar Ecliptic (GSE) coordinate system. By current, we imply current density throughout the text.


\section{Modified Amp\`ere-Maxwell law} 

The dBP equations differ from Maxwell's for the electric field $\vec {E}$ divergence and the  magnetic field $\vec {B}$ curl 
\cite{debroglie-1936}. The latter is given by 

\beq
\frac{\nabla \times \vec{B}}{\mu_0} = \vec{j} + \epsilon_0 \frac{\partial \vec {E}}{\partial t} \underbrace{- \frac{m_\gamma^2 c^2}{\mu_0 \hbar^2}\vec{A}}_{j_{\rm{nM}}}~. 
\label{amperemaxwellmod-1}
\eeq
For $m_\gamma c/{\hbar} = 1/ \lambdabar$, 
$m_\gamma$ is the photon mass, 
$\hbar$ the reduced Planck constant, 
$\epsilon_0$ the permittivity, 
$\mu_0$ the permeability, 
$\vec{j}$ the current density vector, and 
$\vec{A}$ the vector potential.

In the SME framework, the AM law becomes \cite{bonetti-dossantosfilho-helayelneto-spallicci-2017,bonetti-dossantosfilho-helayelneto-spallicci-2018}
 
\beq
\frac{\nabla \times \vec{B}}{\mu_0} = \vec{j} + \epsilon_0 \frac{\partial \vec {E}}{\partial t} \underbrace{+ 
\frac{k^{\rm AF}_{0}\vec{B}}{\mu_0} - \vec{k}^{\rm AF}\times\frac{\vec{E}}{\mu_0 c}}_{j_{\rm{nM}}}~. 
\label{amperemaxwellmod-2}
\eeq 

An effective photon mass may arise if the perturbation vector is purely space-like. Indeed, if $k^{\rm AF}_{0} = 0$, we established \cite{bonetti-dossantosfilho-helayelneto-spallicci-2017,bonetti-dossantosfilho-helayelneto-spallicci-2018} that 

\beq
m_\gamma = \frac{\hbar|{\vec k}^{\rm AF}|}{c}\theta~,
\label{LSVmass}
\eeq  
where $\theta$ is an angular factor depending on the difference between the preferred frame and observer directions 
(for $k^{\rm AF}_{0} = 0$, $\theta$ takes values between $1/4$ and $1$ 
\cite{bonetti-dossantosfilho-helayelneto-spallicci-2018}). 

In our experimental set-up, we don't assess the upper limit of LSV directly, but indirectly: first we assess the mass, Eq. (\ref{photon_mass}), and after the LSV upper limit through the relation, Eq. (\ref{LSVmass}), of mass with the perturbation vector.

%

For NLEM theories, we have set a generalised Lagrangian as a polynomial function of integer powers of the field and its dual:
${\cal L} = {\cal L} ({\cal F},{\cal G})$, 
where 
\begin{align}
{\cal F} = {\ds \frac{1}{2\mu_0}} \left({\ds \frac{{\vec { E}^2}}{c^2}} - \vec { B}^2\right)~~~~~~~~~~~~
{\cal G} = {\ds \frac{1}{\mu_0}}{\ds \frac {{\vec {E}}}{c}}\cdot{\vec {B}}~.
\nonumber
\end{align}
The modified AM equation becomes 
\beq
\frac{\nabla}{\mu_0} \times \left (\frac{\partial \cal L}{\partial \cal F}{\vec B} - \frac{1}{c}\frac{\partial \cal L}{\partial \cal G}{\vec E}\right) =
\vec{j} + \frac{\epsilon_0}{\partial t}\left (\frac{\partial \cal L}{\partial \cal F}{\vec E} + c \frac{\partial \cal L}{\partial \cal G}{\vec B}\right)~.
\label{amperemaxwellmod-3}
\eeq
  
Regardless of the ETEM theory, we can cast Eqs. (\ref{amperemaxwellmod-1},\ref{amperemaxwellmod-2},\ref{amperemaxwellmod-3}) as
\begin{align}
{\vec j}_{B} = {\vec j}_{P} + {\vec j}_{E} + {\vec j}_{nM}~,  
\label{amperemaxwellmod-123}
\end{align}
where ${\vec j}_{B} = \nabla \times \vec{B}/\mu_0$, ${\vec j}_{P} = \vec{j}$, ${\vec j}_{E} = \epsilon_0 
(\partial \vec {E}/{\partial t})$, and ${\vec j}_{nM}$ indicates the non-Maxwellian terms. The latters are pointed in Eq. (\ref{amperemaxwellmod-1}) for the dBP theory, in Eq. (\ref{amperemaxwellmod-2}) for the SME, and, for NLEM theories, by the difference of Eq. (\ref{amperemaxwellmod-3}) with the AM classic law. From Eq. (\ref{amperemaxwellmod-123}), we observe that, if ${\vec j}_{E}$ can be neglected, the difference (scalar or vectorial) between ${\vec j}_{B}$ and ${\vec j}_{P}$ may be attributed to errors or to the non-Maxwellian term ${\vec j}_{nM}$.

Finally, this work focuses on the measurement of current differences in the AM law. The issue of damping and discriminating among theories will be relevant only at the later stage when interpreting the AM law deviations, after experimental errors will be definitively excluded. 


\section{Upper limits on LSV parameters and photon mass}

A disparity from 18 to 24 orders of magnitude opposes the assessments of LSV upper limits:  $|{\vec k}^{\rm AF}| <  4 \times 10^{-7}$  m$^{-1}$ or $5\times 10^{-4}$ m$^{-1}$, and $k^{\rm AF}_0 <  5 \times 10^{-10}$ m$^{-1}$ for laboratory tests \cite{gomesmalta2016}; $|{\vec k}^{\rm AF}| \simeq k^{\rm AF}_0 < 5 \times 10^{-28}$ m$^{-1}$ for astrophysical estimates \cite{kosteleckyrussell2011}. The latter are close to the Heisenberg limit.

Recently, many studies have been conducted on Fast Radio Bursts (FRB), e.g., \cite{boelmasasgsp2016,wuetal2016b,boelmasasgsp2017,shao-zang-2017,wei-zhang-zhang-wu-2017,yang-zhang-2017,wei-wu-2018,xing-etal-2019,Landim-2020,wei-wu-2020,wang-miao-shao-2021,wei-wu-2021,Lin-Tang-Zou-2023,Wang-Wei-Wu-LopezCorredoira-2023},  to set upper limits through dispersion analysis 
\cite{debroglie-1924}, achieving $3.9 \times 10^{-51}$ kg at $95\%$ confidence level \cite{wang-miao-shao-2021}. These estimates may suffer from ambiguities, since the plasma and photon time delays are indistinguishable \cite{bebosp2017}, unless working at gamma-ray frequencies \cite{Bartlett-etal-2021}, or at cosmological distances \cite{boelmasasgsp2017}. 
A space mission was proposed for improving dispersion limits \cite{bebosp2017}. 

The best laboratory test on Coulomb's law sets the limit at $2\times 10^{-50}$ kg \cite{wifahi71}. AM law was tested on ground \cite {Chernikov-etal-1992} and for Cluster \cite{retino-spallicci-vaivads-2016}, where three upper limits were found ranging from $1.4 \times 10^{-49}$ to $3.4 \times 10^{-51}$ kg depending on the value of the potential.

Despite one century of painstaking efforts, we still deal with estimates close to the first value \cite{debroglie-1923}. We consider all explorations around and below $10^{-51}$ kg threshold worth of a scrupulous analysis. 
In the Appendix (\ref{upper}), the Table III \ref{state-of-the-art} 
and Fig. \ref{fig8},
list and display along the years, photon mass upper limits equal and beyond $10^{-50}$ kg. They include the Particle Data Group (PDG) official value of $2\times 10^{-54}$ kg \cite{Ryutov-2007,Workman-etal-2022} and other model-dependent results, some newer than those in the reviews \cite{tulugi05,acciolynetoscatena2010a,scharffgoldhaber-nieto-2010,spqigiro11}.


\section{Method and data analysis} 

For our analysis, we have used the MMS data \footnote{https://mms.gsfc.nasa.gov/index.html}\textsuperscript{,}\footnote{https://lasp.colorado.edu/mms/sdc/public}, contained in the Automated Multi-Dataset Analysis (AMDA), an on-line analysis tool for heliospheric and planetary plasma \cite{genot-etal-2021}.  

AMDA provides different data, among which we find the ion (electron) densities $n_i$ ($n_e$), and the velocities ${\vec v}_i$ (${\vec v}_e$). We do not take the neutrality condition for granted and compute the plasma current directly in AMDA as ($q$ is the charge)
\beq
{\vec j}_P = q (n_i {\vec v}_i - n_e {\vec v}_e)~. \label{particle_current}
\eeq 

In Cartesian components ($x,y,z$), $j_{P}^x = q (n_i v_i^x - n_e v_e^x)$ and similarly for $y,z$; the modulus is $j_P = \sqrt{(j_P^x)^2 + (j_P^y)^2 + (j_P^z)^2}$. 
We computed the error on $j_P$ through propagation from the errors on the velocities and densities provided by AMDA \cite{Gershman-etal-2015,Gershman-etal-2019}, and the average of $j_P$ for the four spacecraft, considering their errors.
For  
$\Delta j_P^x \!=\! q \left(v_i^x \Delta n_i + n_i \Delta v_i^x + v_e^x \Delta n_e + n_e \Delta v_e^x \right)$ and similarly for $y,z$, the total error is 
\beq
\Delta j_P = \frac{1}{j_P}\left(|j_P^x| \Delta j_P^x + |j_P^y| \Delta j_P^y + |j_P^z| \Delta j_P^z\right)~.
\label{error-jp}
\eeq

AMDA provides $j_B$, but not its error. Therefore, we have implemented the curlmeter technique 
\cite{Dunlop-etal-1988,Dunlop-etal-2002} for computing both quantities. We consider a linear approximation of the gradient of the magnetic field measured by the four spacecraft ($\ell=1,2,3,4$). Under this assumption ${\vec j}_B$ is \cite{chanteur-mottez-1993}

\begin{align}
{\vec j_B} = \frac{1}{\mu_0} \sum_{\ell=1}^{4} {\vec k}_\ell \times {\vec B}_\ell~, \label{rotational_current}
\end{align}
where ${\vec k}_l$ are the reciprocal vectors of the tetrahedron, defined by (${\vec r}_{ij} = {\vec r}_j - {\vec r}_i$, where $i$ and $j$ refer to the spacecraft positions) 

\begin{align}
{\vec k}_1 = & \frac{{\vec r}_{23}\times {\vec r}_{24}}{{\vec r}_{21}\cdot({\vec r}_{23}\times {\vec r}_{24})}, \quad
{\vec k}_2 = \frac{{\vec r}_{31}\times {\vec r}_{34}}{{\vec r}_{32}\cdot({\vec r}_{31}\times {\vec r}_{34})}, \nonumber \\
{\vec k}_3 = & \frac{{\vec r}_{24}\times {\vec r}_{21}}{{\vec r}_{23}\cdot({\vec r}_{24}\times {\vec r}_{21})}, \quad
{\vec k}_4 = \frac{{\vec r}_{32}\times {\vec r}_{31}}{{\vec r}_{34}\cdot({\vec r}_{32}\times {\vec r}_{31})} \nonumber ~.
\label{reciprocal_vectors}
\end{align}  

The values of ${\vec j}_B$ found in AMDA compare satisfactorily with our computations. The errors on ${\vec j}_B$ depend on the uncertainties of the magnetic field and on the separations between the spacecraft. For the former, we consider a constant value $\Delta B = 0.1 \times 10^{-9}$ T for each component of the magnetic field \cite{torbert-etal-2016}. This implies an overall error of $0.17$ nT on the modulus of the magnetic field used in our computations. The fluxgate magnetometer offset determination is achieved after two days of solar wind data \cite{plaschke-2019}; for the latter, we consider a relative error equal to $1\%$ of the separation distance \cite{Kelbel-etal-2003}. From the error propagation, we have found that the error on the $q$ component for a single spacecraft reads as 

\begin{align}
    \Delta j_{\ell~B}^q =(|k_\ell^r|+|k_\ell^s|)\Delta B + |B_\ell^r|\Delta k_\ell^s+|B_\ell^s|\Delta k_\ell^r~,
\end{align}
$r,s$ being the other components. The final error on the 
$q\textsuperscript{th}$ component of ${\vec j}_B^q$ is 

\begin{align}
    \Delta j^{q}_B=\sum_{\ell=1}^4 \Delta j_{\ell~B}^q~.
\label{error-jb}
\end{align}

The downloaded data were sampled at $1$ for gathering an extended sample while avoiding high frequency contributions. Since the instruments were polled at a higher rate of tens of milliseconds \cite{Pollock-etal-2016,Russell-etal-2016}, we have verified that AMDA delivers the average of the values belonging to our sampling time, and thus does not pick up a single value each second. Therefore, the high frequency contributions from ${\vec j}_P$ are cut off, allowing the comparison with the low frequency ${\vec j}_B$. We have spanned almost six years of measurements, from November 2015 to September 2021, considering only "burst mode" data \cite{Fuselier-etal-2016} - data with the highest time resolution collected by MMS. Within this data set, we selected the time intervals 
\begin{itemize}
  \item {in which both ${\vec j}_B$ and ${\vec j}_P$ are available,}
  \item {where the four ${\vec j}_P$ currents, measured by each spacecraft (including the error bands) overlap, assuring overlapping of the four outputs.}
 \end{itemize} 
This eases the comparison with ${\vec j}_B$, supposed to be uniform inside the tetrahedron volume, drawn by the four satellites. We have analysed approximately $3.8 \times 10^6$ data points, for each of which we have collected 82 physical quantities: the ion and electron component velocities for four spacecraft (24), distances in components between spacecraft (15), barycentre coordinates (3), the electric field at the first spacecraft (3), the electron and ion densities (8), the parallel and perpendicular electron and ion temperatures (16), the magnetic field (12), and the detection time (1).

\begin{figure} 
    \centering
    \includegraphics[width=1.0\hsize,height=0.3\textwidth,angle=0,clip]{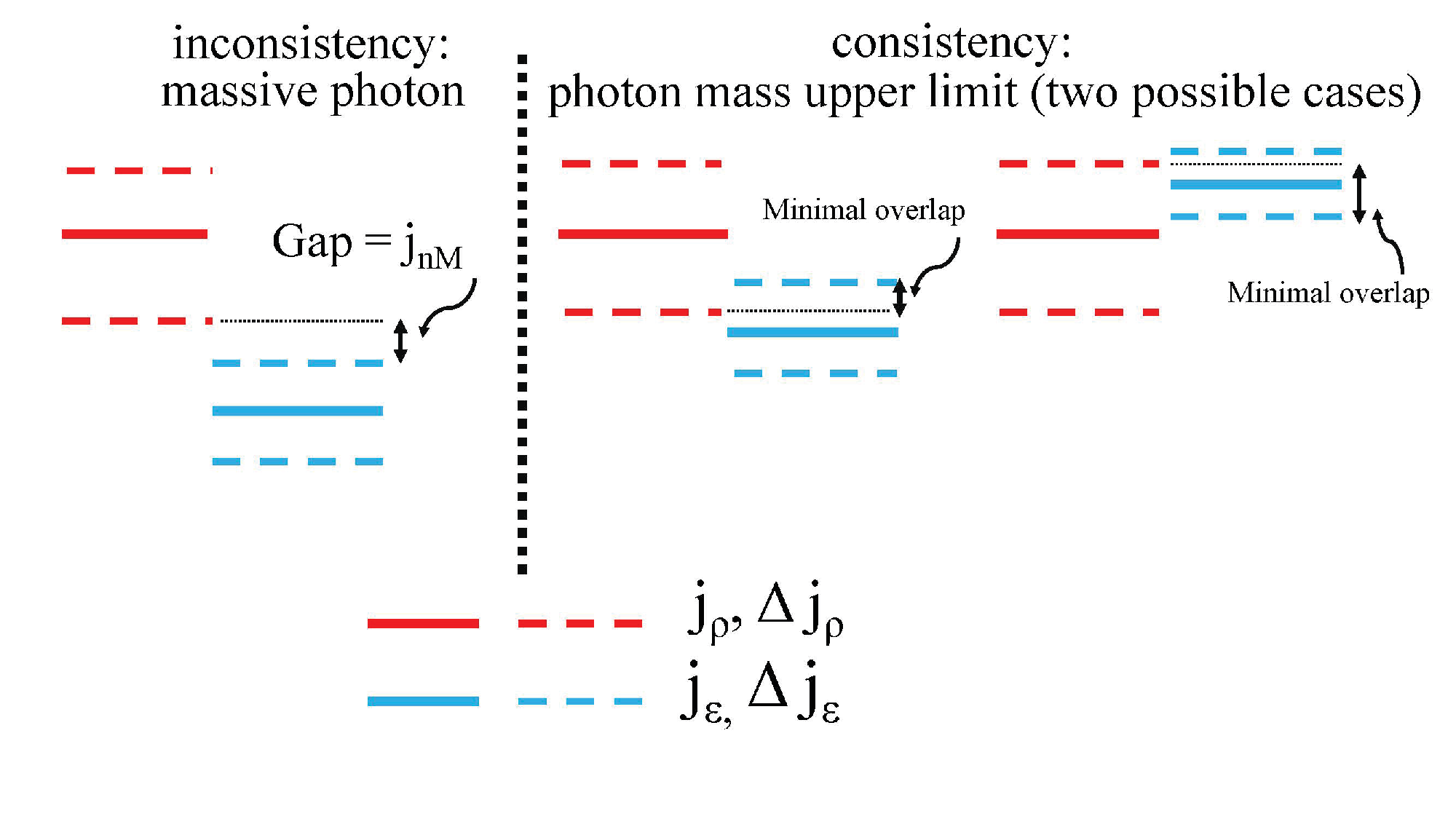}
    \caption{\footnotesize The red (black) and blue (grey) lines represent the currents, and, if dotted, the error bounds. On the left, the case of inconsistency, where the gap implies ${\vec j}_{nM}\neq 0$. On the right, the case of consistency, for which moving upward or downward one of the two bands, we would fall in the inconsistency case, finding just an upper limit for ${\vec j}_{nM}$. This computation has been carried out for the modulus and for each component.}
    \label{fig1}
\end{figure}

The displacement current is computed from the ratio of the electric field variation over the sample interval of 1 s. The average of ${\vec j}_E = 1.4 \times 10^{-14}$ Am$^{-2}$ is six orders of magnitude smaller than the averages of the curl and particle currents and two orders below their smallest difference, Tab. \ref{averages-currents}. This assures us that its contribution is negligible with respect to the other two currents. We thus compared, at each second, ${\vec j}_B$ with ${\vec j}_P$, Eq. (\ref{amperemaxwellmod-123}). We label the comparison ''inconsistency'' if there is a gap between the two current bands - nominal values plus/minus the errors from Eqs. (\ref{error-jp},\ref{error-jb}). We stress that these errors are experimental, obtained from the propagation of the reported instrumental uncertainties. According to our approach, we the gap might be attributed to a non-Maxwellian current ${\vec j}_{nM}$, Eq. (\ref{amperemaxwellmod-123}) or to an undetected error. A ''consistency'' occurs when the two current bands overlap. In these cases, we determine an upper limit to ${\vec j}_{nM}$, as the smallest amount to add (subtract) to (from) one of the two currents to arrive at an inconsistency case, Fig. \ref{fig1}. The procedure is 

\beq
|{\vec j}_{\rho}| - \Delta |{\vec j}_{\rho}| - \left(|{\vec j}_{\epsilon}| + \Delta |{\vec j}_{\epsilon}| \right) = \Delta {j} \underset{overlap}{\overset{gap}{\gtrless}} 0~,  
\label{epsilon}
\eeq
where $\rho,\epsilon = B, P$ depending on whether $|{\vec j}_{P}|- \Delta |{\vec j}_{P}|$ is larger or smaller than $|{\vec j}_{B}|- \Delta |{\vec j}_{B}|$. 

This analysis has been carried out for both the modulus and the Cartesian components. In the latter case, an inconsistency is declared if a gap emerges from at least one of the three axes.

\begin{figure} 
    \centering
    \includegraphics[width=1.0\hsize,height=0.3\textwidth,angle=0,clip]{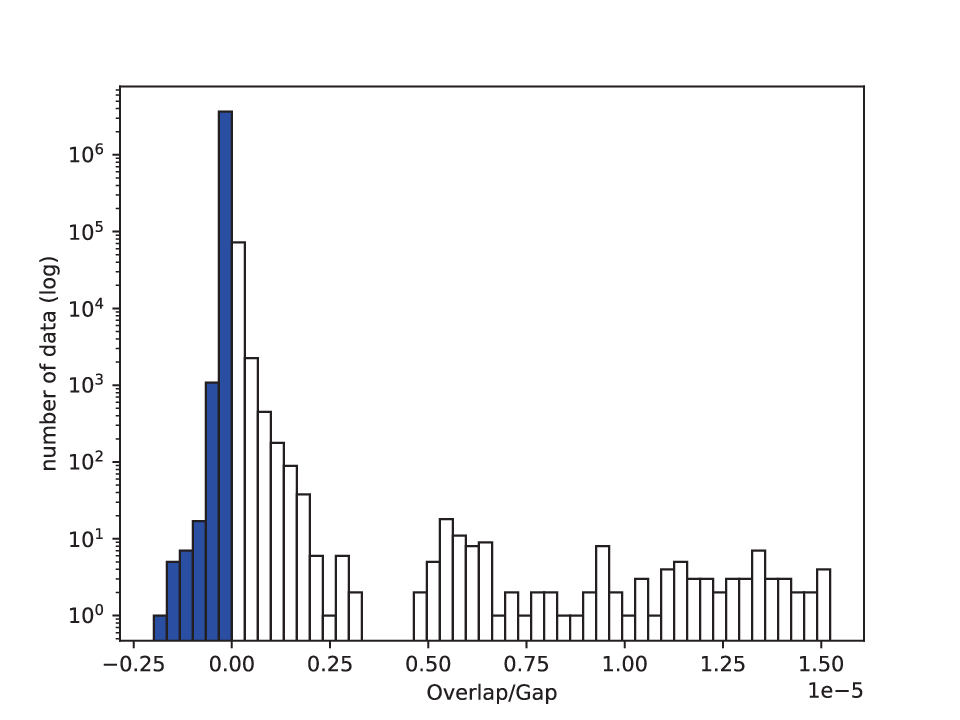}
    \caption{\footnotesize Histogram of Eq. (\ref{epsilon}) for burst data. Most values are negative (overlaps) in blue (grey) bars, on the left of, and close to, zero. The positive values (gaps), white bars, are on the right of zero, and range from $1.1 \times 10^{-12}$ Am$^{-2}$
    to $1.5 \times 10^{-5}$ Am$^{-2}$.}
    \label{fig2}
\end{figure}

Scrutinising the entire data set, we find $2\%$ of inconsistencies for the modulus and $5.2\%$ in Cartesian components; indeed, it may occur that the sum of the currents in absolute values is zero, but their vectorial sum is not, Eq. (\ref{amperemaxwellmod-123}). Such a low number of inconsistencies shoild not be interpreted straightforwardly as a negative outcome of a test on physics foundations, but rather as the consequence of a mission crossing mostly turbulent regions of magnetic reconnection. As later shown, the vast majority of the inconsistencies lay in the solar wind which represents only 14\% of the entire data set, Tab. \ref{results}. 
The results of the analysis of Eq. (\ref{epsilon}) are shown in Figs. \ref{fig2}, \ref{fig3}. It must be added that our analysis considers a large data set of $3.8 \times 10^6$ points, a large sample size, allowing us to find inconsistencies that might have remained unseen in previous experiments. 
For the application of the curlmeter technique, the shape of the tetrahedron drawn by the position of the four satellites matters. Since this method is sensitive to the relative separations between the spacecraft \cite{Dunlop-etal-2018}, we select similar and quasi-optimal configurations. Referring to a regular tetrahedron, we employ the geometrical quality factor \cite{Fuselier-etal-2016} on the volume $Q=V_a/V_r$, where $V_a$ is the actual volume of the tetrahedron in a given moment, while $V_r$ is the volume of the regular tetrahedron having as side the average of the separations between the spacecraft. We have computed $Q$ for data points and set $Q>0.7$ as a threshold for the high quality geometrical factor. The percentages of inconsistencies slightly increase to $2.2\%$ for the modulus and slightly decrease to $4.8\%$ for the components. For the current means, see Tab. \ref{averages-currents}.

 \begin{table}
  \resizebox{\columnwidth}{!}  {
    \begin{tabular}{|c|c|c|c|c|c|}
    \hline
           & $j_B$ & $ \Delta j_B$ & $j_P$ & $\Delta j_P$ \\\hline
    Mean (Gaps) 
    & $4.9 \times 10^{-8}$
    & $6.9 \times 10^{-8}$ 
    & $2.0 \times 10^{-7}$
    & $5.4 \times 10^{-8}$  \\\hline
    Mean (Overlaps) 
    & $2.1 \times 10^{-8}$  
    & $1.8 \times 10^{-7}$ 
    & $3.9 \times 10^{-8}$  
    & $2.4 \times 10^{-8}$   \\\hline
    \end{tabular} 
    }
\caption{The average values [Am$^{-2}$] for the currents and their errors, considering the data with $Q>0.7$.} 
\label{averages-currents}
    \end{table}

We have localised burst data for associating inconsistencies to physical regions of different levels of turbulence, to understand if the underlying physics of plasma influences the results and ultimately the number of inconsistencies. Various criteria to differentiate the plasma regions, have been proposed, {\it e.g.}, \cite{rezeaud-belmont-2018}, which in our case has led to an often ambiguous assignment of data to regions. We have thus identified the regions otherwise. We have computed the ram $p_R$, magnetic $p_M$, and thermal $p_T$ pressures at each data point. If $p_R > 10~max(p_M, p_T)$, we identify the region as solar wind; if 
$ 0.3 ~ min(p_M, p_T) < p_R < 4 ~ max(p_M, p_T)$, as magnetosheath; finally, if $ 10~p_R < p_M$, as magnetosphere. The region of undetermined data points between the solar wind and the magnetosheath is named zone I; between the magnetosheath and magnetosphere, zone II, see Tab. \ref{results}. Remarkably, we observe that the gaps (solar wind + zone I), amount to 76\% (42473/55916) in modulus and 65\% (77847/119850) by components lay in 14\% of the total data.  This seems to indicate a strong effect of the plasma environmental conditions. 

 \begin{table}
  \resizebox{\columnwidth}{!}{
    \begin{tabular}{|c|c|c|c|c|c|}
    \hline
    Results & Solar Wind & Zone I & Magnetosheath & Zone II & Magnetosphere \\\hline
    Gaps (M) & $24376$ & $18097$ & $10669$ & 2203 & $571$   \\\hline
    Overlaps (M) & $92190$ & $261103$ & $1033981$ & $356592$ & $707595$    \\\hline
    Gaps (C) & $34813$ & $43034$ & $35422$ & $5421$ & $1160$  \\\hline
    Overlaps (C) & $81753$ & $236166$ & $1009228$ & $353374$ & $707006$  \\\hline
    Data by region/total data & $4.5\%$ & $9.5\%$ & $42.9\%$ & $14.1\%$ & $29\%$  \\\hline
    Gaps by region/total gaps (M) & $39\%$ & $31.1\%$ & $24\%$ &  $4.5\%$ & $1.4\%$  \\\hline
    Gaps/total data by region (M) & $20.8\%$ & $6.4\%$ & $1.0\%$ & $0.6\%$ & $0.1\%$   \\\hline
    Gaps/total data by region (C) & $29.7\%$ & $15.3\%$ & $3.3\%$ & $1.5\%$ & $0.2\%$  \\\hline
    \end{tabular} 
    }
\caption{Results in modulus (M) and components (C) on gaps (inconsistencies) and overlaps (consistencies) for burst data and $Q>0.7$, analysed by region, in number and percentages. }
\label{results}
    \end{table}


\subsection{Reliability of the analysis} 

The computation of $j_P$ in the solar wind is reliable enough to claim inconsistencies for the following considerations: 
\begin{enumerate}
  \item {we have observed the absence of correlations between inconsistencies and particle densities; thus the inconsistencies are not related to systematic effects due to low densities. Indeed, for the solar wind and zone I inconsistencies, the electron density has an average of $22.96$ cm$^{-3}$ and a median of $11.88$ cm$^{-3}$; the ion density has an average of $22.85$ cm$^{-3}$ and a median of $12.1$ cm$^{-3}$; only in $13.6\%$ and $8.6\%$ of cases the electron and ion densities, respectively, are smaller than $5$ cm$^{-3}$; these values appear sufficient for an adequate measurement of $j_P$. For the inconsistencies in all regions, we add that}
\item {$j_P$ is larger than $j_B$ in 95.6\% of the cases; were the uncertainties provided by AMDA not representative of all sorts of errors, the following cases may arise: }
\begin{enumerate}
\item {$j_P$ is underestimated; this case would not matter: if we had we the 'real' $j_P$, its difference with $j_B$ would increase and the gap would be deeper;}
\item {$j_P$ is overestimated; this case may lead to:}
\begin{enumerate}
\item {lower the number of inconsistencies but also reduce the differences between the two currents and thereby the photon mass, for a given potential (a positive result for our study); or} 
\item {cancel inconsistencies, but this case would be statistically compensated by the 2.(a) and 2.(b).i cases;}
\end{enumerate}
\end{enumerate}
\item {we derive a single $j_P$ by averaging the currents of the four spacecraft, only when the four error bands overlap; by this procedure, we reduce the effect of a possible calibration error.}
\end{enumerate}

We stress that we are more interested in assessing the differences between currents rather than evaluating them exactly.


\subsection{The potential} 

The current differences are strictly the outcome of experimental data, but an estimate of the potential is mandatory to deal with the photon mass (upper limit), Eq. (\ref{amperemaxwellmod-1}). We start by reminding that the potential in massive electro-magnetism is a physical quantity, see, {\it e.g.}, \cite{debroglie-1934d,Marcilhacy-1972, Vigier-Marcilhacy-1972}.
We focus on the solar wind for the following reasons: i) most inconsistencies lay in this region; ii) comparison with other existing model-based \cite{Ryutov-1997} and experiment-based limits \cite{retino-spallicci-vaivads-2016}, derived in this region; iii) the determination of the vector potential in the magnetosheath and magnetosphere is computationally cumbersome \cite{Romashets-Vandas-Poedts-2008,Romashets-Poedts-Vandas-2008} and it deserves a separate work. We use the analytical computation, based on the Parker model, of $|{\vec A}|$  in the Coulomb gauge \cite{bieberetal1987} at each data point. The potential, in spherical coordinates and in the Coulomb gauge, is

\begin{equation} \label{Ar}
 A_r=\frac{2b}{3}\biggr[1-\frac{3}{2}x-x\ln{(1+x)} \biggl]~,
\end{equation}

\begin{equation} \label{Atheta}
A_{\theta}=\frac{2b}{3} \sin \theta \biggr[\frac{x}{1+x}+\ln(1+x) \biggr] \biggr (\frac{\cos \theta}{x} \biggl)~,
\end{equation}

\begin{equation} \label{Aphi}
A_{\phi}=\frac{a}{r \sin \theta}(1+x)~,
\end{equation}
where $x=|\cos \theta|$ and, at 1 AU and for magnetic fields equal to 5 \, nT, $a=3.54\times 10^{-9}$ T AU$^2$, $b=3.54\times 10^{-9}$ T AU. Near the Earth for the modulus of the potential, a direct proportionality emerges, that is $A=\sqrt{11/6}~BR$, where $R$ is 1 AU. 

For the adoption of the  Coulomb gauge, we recall that for deriving the dBP equations from the Lagrangian, we impose the Lorenz condition. If the scalar potential is (almost) time independent, then the Coulomb and Lorenz conditions can be considered equivalent.  


\section{Upper limits}

\begin{figure} 
    \centering
    \includegraphics[width=1.0\hsize,height=0.3\textwidth,angle=0,clip]{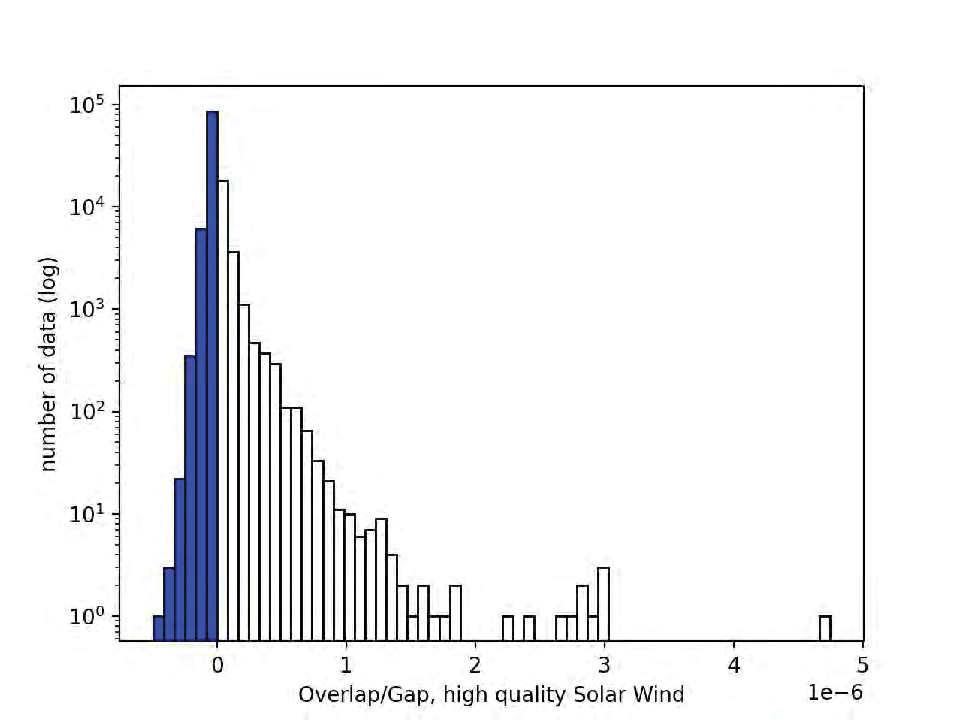}
    \caption{\footnotesize Histogram of Eq. (\ref{epsilon}) for burst data ($Q > 0.7$) in the solar wind. Most values are negative (overlaps) in blue (grey) bars, on the left of, and close to, zero. The positive values (gaps), white bars, are on the right of zero, and range from $2.0 \times 10^{-12}$ Am$^{-2}$ to $4.7 \times 10^{-6}$ Am$^{-2}$.    }
    \label{fig3}
\end{figure}

We can infer an upper bound of the photon mass. The found distributions are strongly asymmetric and thus the standard deviation on the mean measurement would be misleading. 

We treat the data concerning the inconsistencies (gaps) and consistencies (overlaps) in a singular statistical analysis. We restrict our analysis to the values shown in Fig. \ref{fig3}, and use this expression   

\beq
m_\gamma = \frac{\hbar{\sqrt {\mu_0}}}{c}\sqrt{\ds\frac{|\Delta{\vec j}|}{|{\vec A}|}} ~{\rm~kg}~, \label{photon_mass}
\eeq 
which we can compute in the solar wind. Once the mass upper limit is determined through Eq. (\ref{photon_mass}), we derive a LSV upper limit through Eq. (\ref{LSVmass}).    

For the analysis of the entire data set we refer to the definitions in Fig. \ref{fig1}, recalling that the consistencies are defined as negative, while the inconsistencies are positive. From Fig. \ref{fig3}, we compute the median and the $25 \%$ and $75 \%$ percentiles, in order to find a proper confidence interval for our results. 

We consider the modulus of the consistencies to assess the upper bound of the photon mass from our analysis of the entire set of data. We find a median value equal to $2.7 \times 10^{-8}$ Am$^{-2}$, with a $25 \%$ percentile equal to $7.9 \times 10^{-9}$ Am$^{-2}$ and a $75 \%$ percentile of $4.7 \times 10^{-8}$ Am$^{-2}$.

Having computed also the median for the potential in the solar wind data, $1.8 \times 10^3$ T m, we derive an estimate equal to $2.1 \times 10^{-51}$ kg, with a confidence interval going from $1.1 \times 10^{-51}$ kg to $2.8 \times 10^{-51}$ kg. This upper limit levels to the recent observational estimates derived by FRB. 

For the LSV parameter $|\vec{k}^{\rm AF}|$, Eq. (\ref{LSVmass}), the median is $6 \times 10^{-9}$ m$^{-1}$, for $\theta = 1$, in line with laboratory upper limits.


\section{The question of the inconsistencies in the solar wind}

After having found the upper limits, we now focus on the inconsistencies jn the solar wind.  

In this section, we suppose that the inconsistencies (gaps) in the solar wind are manifestations of non-standard electro-magnetic effects, Eq. (\ref{photon_mass}), and not of unaccounted experimental errors. In other words, we reverse the ordinary thinking and see if the found mass value is compatible with existing upper limits. The objective is not claiming a photon mass discovery, but rather emphasising that the evidence of the masslessness of the photon at the $10^{-53}$ kg level is worthy a scrupolous analysis. Therefore, we proceed by i) addressing the solar wind, ii) analysing the official limit in the literature, iii) examining our results in the solar wind regime.   

\paragraph{ The nature of the solar wind.} We refer to the nearly collisionless nature of the solar wind implying that the mean free path of the particles is in of the order of 1 AU \cite{Sahraoui-etal-2020}. Thereby,  the kinetic energies and momenta of the particles change only as a result of the averaged fields generated by the other particles. How turbulence explains the acceleration of plasma particles, {\it i.e.} the transfer of energy across a broad range of scales that leads to complex chaotic motions, structure formation and energy conversion, is a marginally pertinent for this work. 
Indeed, the solar wind flows rapidly towards Earth carrying the solar magnetic field. When it smacks right into the terrestrial magnetic field, it generates the bow shock and the current flow becomes turbulent. Until then, the solar wind is possibly the least troublesome region for testing the AM law. Thus, the region of the solar wind before the shock presents the best conditions to test the AM law.     

\paragraph {Comparison with the literature.} We discuss the PDG \cite{Workman-etal-2022} adopted limits coming out of modelling the solar wind magnetohydro-dynamics at 1 AU \cite{Ryutov-1997} and later at 40 AU \cite{Ryutov-2007}. 
In \cite{Ryutov-1997}, the extensive application of the Parker model  \cite{Parker-1958} couples to the absence of data and of error analysis. The final estimate is based on reconciling theory with the non-observation of large plasma motions in the solar wind. There is not an actual upper limit being stated, but a supposed improvement of a factor of 10 with reference to a differently obtained estimate \cite{Davis-Goldhaber-Nieto-1975} which concerns Jupiter data at 5.2 AU.

The $1.4 \times 10^{-54}$ kg limit \cite{Ryutov-2007}
stands as the actual official upper limit. As the former, it is based entirely on the Parker model, on a qualitative reasoning {\it ad absurdum}, while data are not presented and an error analysis is absent; further,the estimated feebleness of the ${\vec j} \times {\vec B}$ force and the absence of deceleration of the radial expansion at Pluto orbit. But recent results \cite{Elliott-etal-2019} have conversely shown that the solar wind slows down reaching Pluto.  The final estimate,  Eq. (14) in \cite{Ryutov-2007}, comes from a single set of three data: magnetic field, ion density and particle velocity, taken from the missions Pioneer and Voyager; if these values were to vary of just $70\%$ - as it may easily happen - the limit would worsen of one order of magnitude. For concluding this part, we refer to a more recent model analysing rigorously the implications of the non-zero photon mass on ${\vec j} \times {\vec B}$. From this analysis, it appears that the force-free equilibrium postulated by Ryutov is rather unlikely \cite{Bhattacharjee-2023}.  

Our comments aim solely to show the shortcoming of a legitimate, but model-based upper limit. 

\begin{figure}
    \centering
    \includegraphics[width=1.0\hsize,height=0.3\textwidth,angle=0,clip]{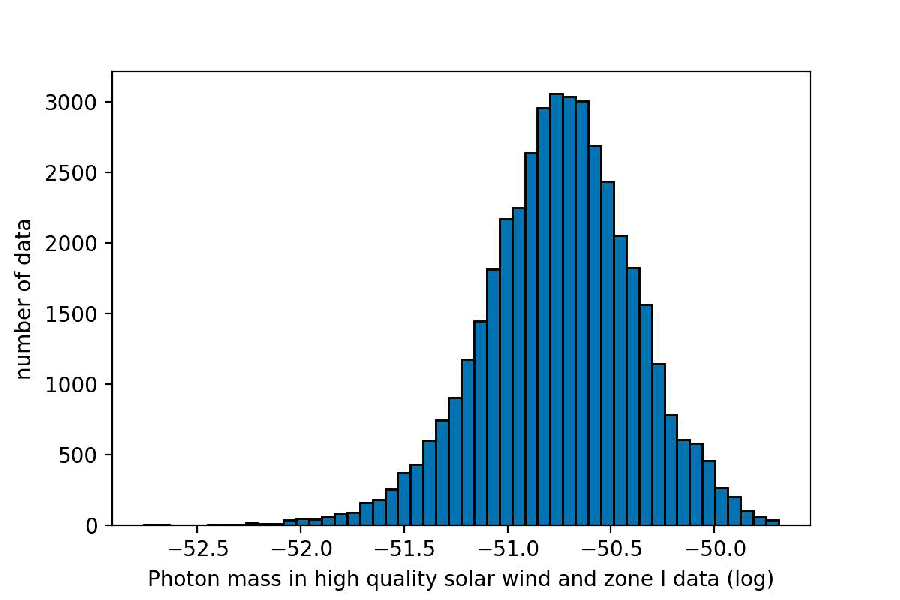}
        \caption{\footnotesize For the solar wind and zone I regions burst data and $Q>0.7$, the histogram of the dBP photon mass derived from the inconsistencies, supposing that the latter are not associated with unaccounted errors. The the median of the distribution is $2.5 \times 10^{-51}$ kg, which is a close value to the previously presented upper limit.} 
    \label{fig4}
\end{figure}

\begin{figure}
    \centering
    \includegraphics[width=1.0\hsize,height=0.3\textwidth,angle=0,clip]{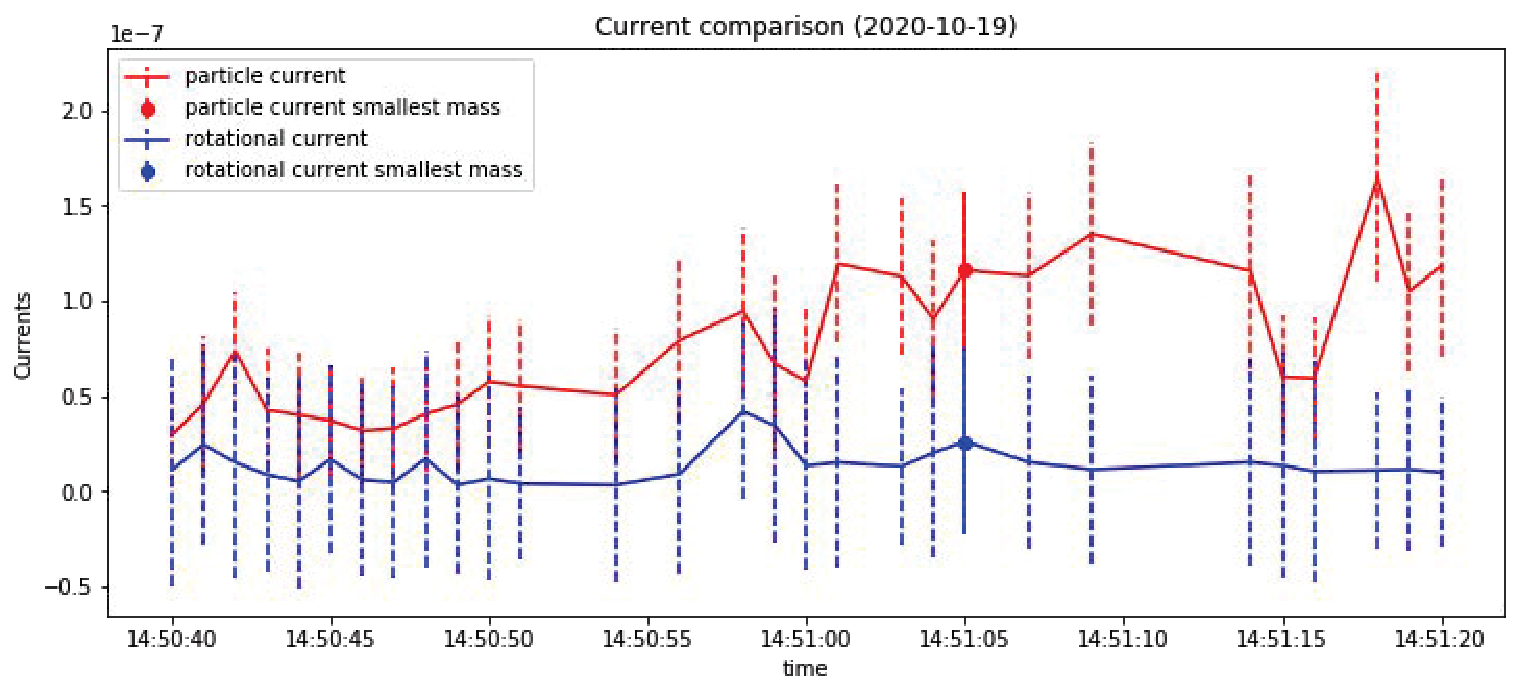}
    \caption{\footnotesize Supposing that all errors have been considered, the smallest dBP photon mass, see the spots, occurs on 19 October 2020, at 14h 51m 05s, in the zone I burst data and $Q = 0.82$. The particle current $j_P = (1.2 \pm 0.4 ) \times 10^{-7}$ Am$^{-2}$, and the rotational current $j_B = (2.6 \pm 4.9) \times 10^{-8}$ Am$^{-2}$ produced a difference of $7.4 \times 10^{-12}$ Am$^{-2}$. This difference is not visible on the vertical axis carrying units which are five orders of magnitude larger. The vertical dotted lines are the errors. The potential of $3.8 \times 10^3$ T m leads to a mass of $1.7 \times 10^{-53}$ kg, Eq. (\ref{photon_mass}), and thereby $|\vec{k}^{\rm AF}| = 5 \times 10^{-11}$ m$^{-1}$, Eq. (\ref{LSVmass}).}
    \label{fig5}
\end{figure}

\begin{figure}
    \centering
    \includegraphics[width=1.1\hsize,height=0.33\textwidth,angle=0,clip]{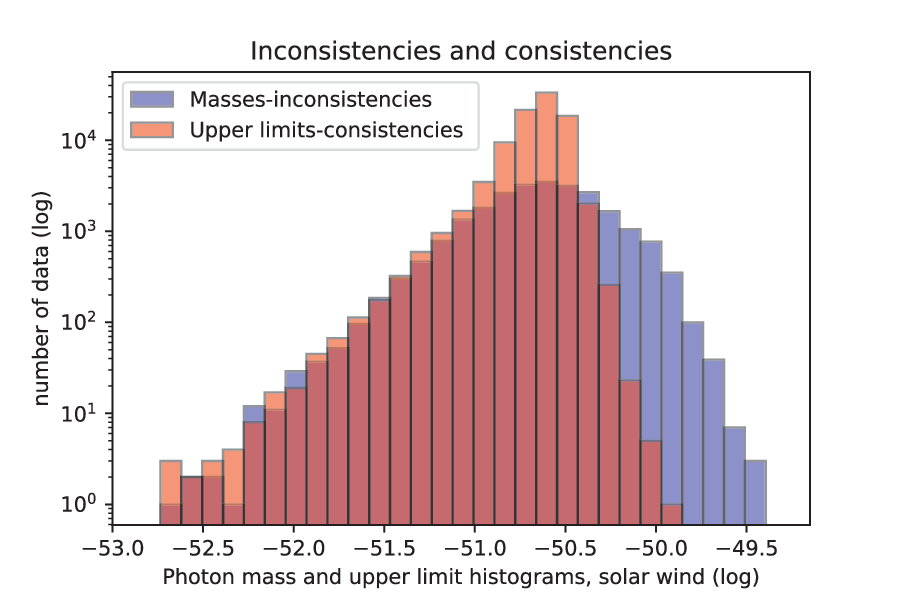}
        \caption{\footnotesize For the solar wind region burst data and $Q>0.7$, the histograms of the dBP photon mass upper limits derived from the consistencies and the dBP photon mass values derived from the inconsistencies, supposing that the latter are not associated to unaccounted errors. Only the inconsistencies in absolute value are shown, representing 20.8\% of the total data in the region. We recall that the inconsistencies in components are 29.7\% in the region.} 
    \label{fig6}
\end{figure}

\begin{figure}
    \centering
    \includegraphics[width=1.1\hsize,height=0.33\textwidth,angle=0,clip]{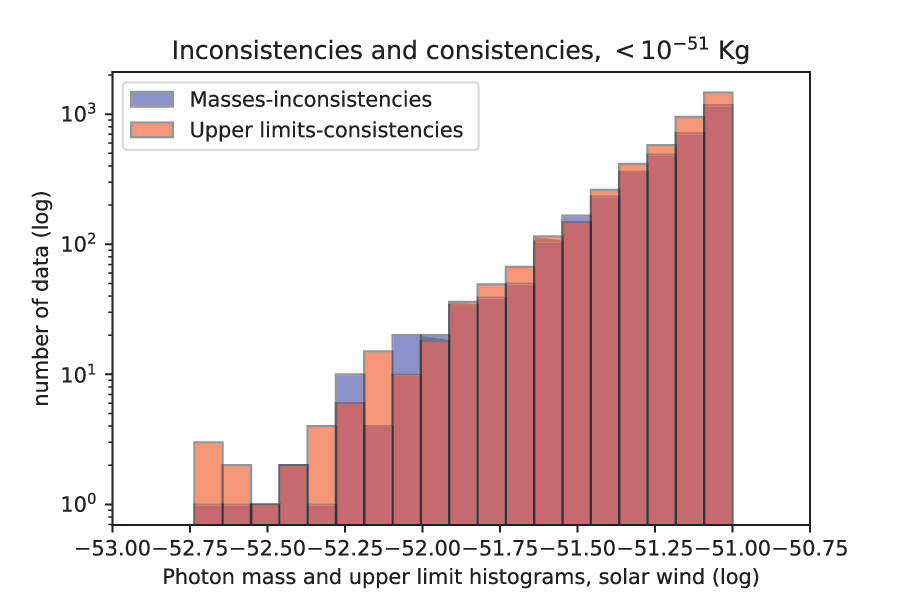}
        \caption{\footnotesize Same caption of the previous figure. The values below $10^{-51}$ kg are zoomed in.} 
    \label{fig7}
\end{figure}

\paragraph{Deviations from the Amp\`ere-Maxwell law.} 

Figure \ref {fig4}, displays the dBP photon mass estimates from the inconsistencies. We learn that: i) the median for solar wind and zone I is $1.8 \times 10^{-51}$ kg, which is close to the value of the photon mass upper limit; ii) if we mean by minimum the mass value that would fit all inconsistency cases, this would be $1.7 \times 10^{-53}$ kg. 

Incidentally, since the potential is space-time dependent, the minimum photon mass does not necessarily correspond to the minimum current deviation, $7.4 \times 10^{-12}$ Am$^{-2}$, Fig. \ref {fig5}. 

In Figs. \ref{fig6}, \ref{fig7}, we show the two distributions of overlaps and gaps. They look very similar, due to our definition of the upper limit, involving the minimum displacement, upward or downward, of the current bands, 
Fig. \ref{fig1}. Indeed, the two distributions almost overlap, especially for values of mass upper limits and the dBP photon mass below $10^{-51}$ kg, Fig. \ref{fig7}. In this domain the percentages of inconsistencies arrive up to 45.2\%. 

The similarity of the distributions for weak current differences in the solar wind warns against all attempts of using oversimplified models.

Table III shows that around $10^{-53}$ kg level and beyond, there is crowding of model-based limits. 


\section{Discussion, conclusions, perspectives} 

Considering the entire set of consistencies and inconsistencies in the solar wind, we found an estimate on the upper bound for the photon mass equal to $2.1 \times 10^{-51}$ kg, and for the LSV parameter $|\vec{k}^{\rm AF}|$, which is $6 \times 10^{-9}$ m$^{-1}$. The photon mass upper limit is consistent with various other upper limits found from recent FRB observations, Tab. III, while the LSV upper limit is compliant with terrestrial experiments \cite{gomesmalta2016}. 

After error analysis, we have found deviations from the AM law in 2.2\% of the cases for the modulus and 4.8\% for the vector components.  
Such a paucity of cases turns into a large minority if the burst modes are counted only in the solar wind. We then find inconsistencies in 20.8\% of the cases for the modulus and 29.7\% for the components. 

The identification of numerous deviations from the AM law might be related to our analysis encompassing a large data set of $3.8 \times 10^6$ points, allowing us to uncover inconsistencies that might have remained undetected in earlier studies. 

We lack {\it experimental} constraints to {\it definitively} rule out a mass at $10^{-53}$ level, Tab. III. Many published limits are identified as speculative \cite{Workman-etal-2022}. Ref. \cite{scharffgoldhaber-nieto-2010} states
''Quoted photon-mass limits have at times been overly optimistic in the strengths of their characterizations. This is perhaps due to the temptation to assert too strongly something one {\it knows} to be true''. 

Although we find deviations from Amp\`ere-Maxwell law for some data points, we do not intend to claim it as the final result. We cannot exclude possible mismatches between currents, even after instrument calibration and analysis of systematic effects on MMS \cite{Gershman-etal-2015,Fuselier-etal-2016,Pollock-etal-2016,Russell-etal-2016,Gershman-etal-2019}, as the intrinsic charge \cite{Barrie-et-al-2019}. Reducing systematic errors to a minimum is an objective for MMS. The instrumentation enables regular cross-calibration and validation of the Field Plasma Investigation (FPI) data, reducing systematic errors to within a few percent, thus providing suitable accuracy to calculate the current directly from particle observations \cite{Gershman-etal-2019}. For the Digital Flux Gate (DFG) and the Analogue Flux Gate (AFG) magnetometers, the ground calibration is refined in space to ensure all eight magnetometers are precisely inter-calibrated \cite{Russell-etal-2016}. 

The values of the upper limit of distribution are similar to the values of the photon mass inferred from the inconsistencies distribution due to our rigorous definition of the upper limit, involving the minimum displacement, upward or downward, of the current bands.

Summarising, our main result is not the upper limit on the photon mass, nor do we claim to have found a massive photon. Our main result is having found numerous inconsistencies (up to 30\%) in the best sub-set of data (in the solar wind, which is the best for our investigation since it possibly corresponds to the least turbulent region \cite{Neugebauer-1976} crossed by MMS). These inconsistencies are likely due to unknown experimental errors, but we cannot exclude non-Maxwellian terms. The preceding should be seen in the context of a non-dedicated mission, since the aforementioned 'best data' set is only five percent of the 3.8 million data. Still, even if we consider only this subset, the number of data analysed is enormous when compared to qualitative assessments based on models and on one single set of values.

The determination of the nature of the deviations is a challenge that requires exploring current differences well below $10^{-11}$ Am$^{-2}$, Eq. (\ref{photon_mass}), ideally four orders of magnitude, to reach the $10^{-55}$ kg area. While this is ambitious, it paves the way to new and far reaching, combined explorations in plasma and fundamental physics, which are usually separated domains of research.
The accuracy in measuring the velocity difference of ions and electrons $v_i - v_e = (\Delta j + j_B)/qn$ in the order of $1.5 \times 10^4$ ms$^{-1}$ is technologically feasible, conversely to Cluster, where $j_B\ll j_P$ led to $10^{-4}$ ms$^{-1}$ as required velocity difference \cite{retino-spallicci-vaivads-2016}.
Since the curlmeter method does not allow to estimate currents in regions smaller than the scale of the tetrahedron itself, a constellation of more than four spacecraft targeting the solar wind with high quality data would be desirable.
Indeed, future satellite measurements may clarify the nature of the deviations, whether unaccounted errors or profoundly meaningful first glimpses of new physics.
Despite the growing gravitational astronomy, photons remain our main tool for interpreting the Universe. Being herein confronted with a non-dedicated mission to fundamental physics, we urge to not spare efforts to explore the foundations of physics.

\paragraph*{Acknowledgements.} 
CNES and the Federation {\it Calcul Scientifique et Mod\'elisation Orl\'eans Tours} (CaSciModOT) for missions and the student internships of S. Dahani and A. Gauthier are thanked. Discussions with G. Chanteur, H. Breuillard and A. Retin\`o (Paris), V. G\'enot (Toulouse), D. Gershman (NASA), B. Lavraud (Bordeaux), A. Vaivads (Stockholm), E. Vitagliano (Los Angeles) and the colleagues T. Dudok de Wit, P. Henri, V. Krasnosselskikh, J.-L. Pin\c{c}on (Orl\'eans) are  acknowledged. GS thanks the hospitality of LPC2E, the support by the {\it Istituto Nazionale di Fisica Nucleare, Sez. di Napoli, Iniziativa Specifica QGSKY} and the Erasmus+ programme.


\begin{widetext}

\section{Appendix}

\subsection{The Proca conception of the photon \label{proca}}

Proca conjectured that the photon would be composed of two pure charge massless particles, one positive and one negative. The Lagrangian in Proca's original notation is \cite{proca-1936d, proca-1937}
\beq
{\cal L} = h^2 c^2 \left[ \frac{1}{2} F_{rs}G_{rs} + k^2 \psi^*_r\psi_r \right]~,
\eeq
where $F_{rs} = (\partial_r + i A_r)\psi_s^* - (\partial_s + i A_s)\psi_r^*$ and $G_{rs} = (\partial_r - i A_r)\psi_s - (\partial_s - i A_s)\psi_r$ are tensorial fields; $A_s= e\phi_s/hc$, being $e$ the charge, $h$ the Planck constant, $c$ the light speed and $\phi$ the field potential; $\psi$ is a wave-function ({\it vecteurs d'univers)} and $k = mc/h$. Proca affirmed that for photons $\psi^*_r=\psi_r$ and $m=0$. 

\subsection{The question of mass and damping in the SME \label{LSV-SME}}

Frame dependency is unusual for mass, but similarly to the non-covariance of energy in general relativity, we should not be refrained from using the concept of mass. What distinguishes an effective from a physical mass? Massless particles such as charged leptons, quarks, W, and Z Bosons acquire mass through the Higgs mechanism, while composite hadrons (baryons and mesons) from Chiral Symmetry (Dynamical) Breaking (CSB). How to qualify the mass induced by LSV? In Quantum Field Theory (QFT), a physical mass is a quantity calculated
perturbatively by considering loop effects and computing the poles of the loop-corrected propagators. Since the CFJ term respects gauge symmetry and there is no room
for anomalies in the model, loop corrections do not shift the pole of the tree-level propagator. This implies that the mass may be interpreted both effective and physical.

For the CPT-odd handedness, ${k}^{\rm AF}_0$ and $\vec{k}^{\rm AF}$ are the time and space components of ${k}^{\rm AF}$
$ \left({k}^{\rm AF}_0, \vec{k}^{\rm AF}\right)$. For 

\beq
\begin{cases}
& \bar{\omega}= \omega/c~,\\ 
& k^{\mu} = \left(\bar{\omega},\vec{k}\right)~, \\ 
& k^2 = \bar{\omega}^2 - |\vec{k}|^2~, \\ 
& \left(k^{\rm AF}_{\mu} k^{\mu}\right)^{2} = \left(k^{\rm AF}_0 \bar{\omega} - {\vec k}^{\rm AF} \cdot {\vec k} \right)^2~, 
\end{cases}
\eeq
the CFJ dispersion relations at fourth order in the 4-momentum $k$  
\cite{bonetti-dossantosfilho-helayelneto-spallicci-2017,bonetti-dossantosfilho-helayelneto-spallicci-2018}, originally in \cite{carroll-field-jackiw-1990}, are  
  
\begin{eqnarray}
& k^4 + ({k}^{\rm AF})^2 k^2 - ({k}^{\rm AF} \cdot k)^2 = 0 = \nonumber \\
&\left(\bar{\omega}^2 \!-\! \vec{k}^{~2} \right)^2 \!\!+\!
\left[\left({k}^{\rm AF}_0\right)^2 \!\!-\! \left({\vec k}^{\rm AF}\right)^2\right] \left[\bar{\omega}^2 \!-\! \vec{k}^{~2} 
\right] \!-\! 
\left({k}^{\rm AF}_0 \bar{\omega} \!-\! {\vec k}^{\rm AF}\!\cdot {\vec k}\right)^2 
\label{drlsv}
\end{eqnarray}

We consider three cases. 

Case 1. Taking ${\vec k} = 0$, Eq. (\ref{drlsv}) becomes 

\beq
\bar{\omega}^2 \left [\bar{\omega}^2 - ({\vec k}^{\rm AF})^2 \right] = 0~.
\eeq 

The roots are $\bar{\omega} = \left(0, |{\vec k}^{\rm AF}|\right)$, for any value of $k^{\rm AF}_0$. Zero momentum does not forcefully imply zero photon energy, unless ${k}^{\rm AF}$ is purely time-like. But time-likeness causes a problem with the unitarity of the quantum version of this theory \cite{Adam-Klinkhamer-2001}. 

Case 2. We now consider $k^{\rm AF}_0 = 0$. Eq. (\ref{drlsv}) becomes
\beq
\bar{\omega}^4 - \left [\left({\vec k}^{\rm AF}\right)^2 + 2{\vec k}^{~2 }\right]\bar{\omega}^2 
+ \left[\left({\vec k}^{\rm AF}\right)^2 {\vec k}^{~2} - \left({\vec k}^{\rm AF}\cdot {\vec k}\right)^2 + {\vec k}^4\right] = 0~.
\eeq

The roots are $\bar{\omega}^2 = {\vec k}^{~2} + \frac{1}{2}\left({\vec k}^{\rm AF}\right)^2 
\pm \frac{1}{2}\sqrt{\left({\vec k}^{\rm AF}\right)^4 + 4\left({\vec k}^{\rm AF}\cdot {\vec k}\right)^2}$  

Case 3. If as for Case 2, $k^{\rm AF}_0 = 0$ but also ${\vec k}^{\rm AF} \cdot {\vec k} = 0$, the roots simplify to 
 $\bar{\omega}^2 = \left[{\vec k}^{~2}, {\vec k}^{~2} +\left({\vec k}^{\rm AF}\right)^2 \right]$~.
Here, the second set of roots reminds of the dBP dispersion relation.

In addition to the above demonstration, other arguments for the massiveness of the photon can be put forward. Before doing so, we remind that the number of polarisations is depending on our gauge choice which depends on our assumptions and on the experimental evidence. 

\begin{itemize}
\item{For a massless photon, the longitudinal polarisation is forbidden since it implies that the photon would travel at $v<c$ (due to the vibrations in the direction of motion), while in the massive case, there is compatibility.}
\item{The CFJ Lagrangian was rewritten \cite{bonetti-dossantosfilho-helayelneto-spallicci-2017} to display a term 
$|\vec{k}^{\rm AF}|~A_i A^i$, $A_i$ being the four-potential as in the dBP Lagrangian. Electro-dynamics models exhibiting this term have a modified constraint algebra due to the gauge symmetry breaking \cite{Diez-etal-2020}. This causes the propagation of an additional degree of freedom, leading to a total of three.}
\item{As already stated, in \cite{bsbebohn2003,baetaetal2004}, it was shown that a LoSy breaking includes a massive mode in the propagator proportional to the violation vector.}
\item{In {\it vacuo}, while $\nabla {\vec B}$ remains zero, this is not the case anymore for $\nabla {\vec E}$. The energy-momentum
relation that emerges as a dispersion relation neatly points out the presence
of a mass, much in the way of a Klein-Gordon dispersion relation.}
\item{The Higgs mechanism does not forbid the creation of a photon mass, also in the LSV context  \cite{itzykson-zuber-2006,addvgr-2007}.}
\item{ It has been shown \cite{Altschul-2006}, that by computing the radiative correction of the photon self-energy, two masses may arise. In the pure time-like case, photons assume a tachyonic behaviour.}
\item{ In \cite {Altschul-2006} the Fermionic sector induces a photon mass through the LSV. In our work 
\cite{bonetti-dossantosfilho-helayelneto-spallicci-2017,bonetti-dossantosfilho-helayelneto-spallicci-2018}, the CFJ term is already included in the model and we don't need to step back to the Fermionic origins. In this sense, our work 
\cite{bonetti-dossantosfilho-helayelneto-spallicci-2017,bonetti-dossantosfilho-helayelneto-spallicci-2018}
may be considered \cite{Altschul-2006}.}
\end {itemize}

{ In the SME framework, damping is a complex issue. A purely time-like $k^{\rm AF}$ vector was considered \cite{Altschul-2014,Schober-Altschul-2015,Schober-Altschul-2016}, despite the issue of unitarity \cite{Adam-Klinkhamer-2001}. In the last of these references, the divergence of $\vec E$ is zero, whereas the curl of $\vec B$ differs from the corresponding Maxwellian form. Therefore the long-distance behaviour of the electric and magnetic field differs. The same reference states that there are important differences between time-like and space-like cases. 

Indeed, the damping for the space-like case} is the same for both electric $\vec{E}$ and magnetic $\vec{B}$ fields as their Green functions share the same poles, { regardlessly of the kind of source, in the static cases, see below. 
Nevertheless,} even if the damping is identical, the coupling of the perturbation vector with the ${\vec E}$ and ${\vec B}$ fields, in the Gauss and AM laws, determines a direction-dependent damping. 


\subsubsection*{Damping for electric and magnetic fields, for a space-line perturbation vector $k^{\rm AF}$.}

In a static regime, for whatever source and in absence of a current, the wave equations for the potentials $\phi$ and $ \vec{A}$ read as follows

\begin{align}
   & \left[\left(\nabla^2\right)^2-\left({\vec k}^{\rm AF}\right)^2 \nabla^2+\left({\vec k}^{\rm AF}\cdot \vec{\nabla}\right)^2\right]\phi = - \nabla^2 \frac{\rho}{\epsilon_0}-\mu_0 c \left.\right.{\vec k}^{\rm AF}\cdot \left(\Vec{\nabla}\times\Vec{j}\right)&\\
   &\left\{\left[\left(\nabla^2\right)^2-\left({\vec k}^{\rm AF}\right)^2 \nabla^2+\left({\vec k}^{\rm AF}\cdot \vec{\nabla}\right)^2\right]\delta_{ij}+\left({\vec k}^{\rm AF}\right)^2\partial_i \partial_j +k^{\rm AF}_i k^{\rm AF}_j \nabla^2-
\left({\vec k}^{\rm AF}\cdot \vec{\nabla}\right)\left(k^{\rm AF}_i\partial_j+k^{\rm AF}_j \partial_i \right)\right\}A_j= \frac{1}{\epsilon_0 c}\left({\vec k}^{\rm AF}\times \Vec{\nabla}\rho\right)_i- \left(\mu_0\nabla^2 \vec{j}\right )_i
\end{align}


By using the AM law, we get the following relation in Fourier space

\begin{align}
    \vec{ \tilde A} ({\vec k}) = - \frac{1}{c} \frac{1}{k^2}{\vec k}^{\rm AF}\times \vec{\tilde {E}} ({\vec k})
\end{align}

This in turn will give for the electric and magnetic fields the following expressions in the case of a static point-like charge $q$
\begin{align}
   & \phi= \frac{q}{\epsilon_0}\int \frac{d^3 \Vec{k}}{(2\pi)^3}\frac{1}{{\vec{k}}^2+ \left({\vec k}^{\rm AF}\right)^2-
\left({\vec k}^{\rm AF}\cdot\hat{k}\right)^2}e^{-i\Vec{k}\cdot \Vec{x}}& \label{alt1} \\
   &\Vec{E}= - \Vec{\nabla}\phi = \frac{q}{\epsilon_0}\int \frac{d^3 \Vec{k}}{(2\pi)^3}\frac{i\Vec{k}}{{\vec{k}}^2+
\left({\vec k}^{\rm AF}\right)^2-\left({\vec k}^{\rm AF}\cdot\hat{k}\right)^2}e^{-i\Vec{k}\cdot \Vec{x}}& \label{alt2} \\
   &\Vec{B}= \frac{1}{c}\frac{1}{{\nabla}^2}\Vec{\nabla}\times({\vec k}^{\rm AF}\times \vec{E})=\frac{-q}{c \epsilon_0}
\int \frac{d^3 \Vec{k}}{(2\pi)^3}\frac{{\vec k}^{\rm AF}-({\vec k}^{\rm AF}\cdot \hat{k})\hat{k}}{{\vec{k}}^2+
\left({\vec k}^{\rm AF}\right)^2-\left({\vec k}^{\rm AF}\cdot\hat{k}\right)^2}e^{-i\Vec{k}\cdot \Vec{x}}~. \label{alt3}
\end{align}

The damping in the fields is fixed by the poles of the integrands, which share a common denominator. According to Eqs. (\ref{alt2},\ref{alt3}), there are two poles, common to both the integrals, and they are purely imaginary: $i |{\vec k}^{AF}| [\sin \alpha|$ and 
$- i |{\vec k}^{AF}| |\sin \alpha[$, where $\alpha$ is the angle between ${\vec k}^{AF}$ and ${\vec k}$. As usual, these integrals may be calculated in complex plane. Since the ${\vec E}$ and ${\vec B}$ fields should go to zero at infinity, only the pole 
$- i |{\vec k}^{AF}| |\sin \alpha|$ is retained. Thus, for a space-like background, the electro-static and magneto-static fields exhibit the same spatial damping, contrary to what happens for a time-like $k^{AF}_\mu $.





\subsection{Upper limits \label{upper}}

\begin{figure}[h]
  \centering
    \includegraphics[width=1.0\hsize,height=0.5\textwidth]{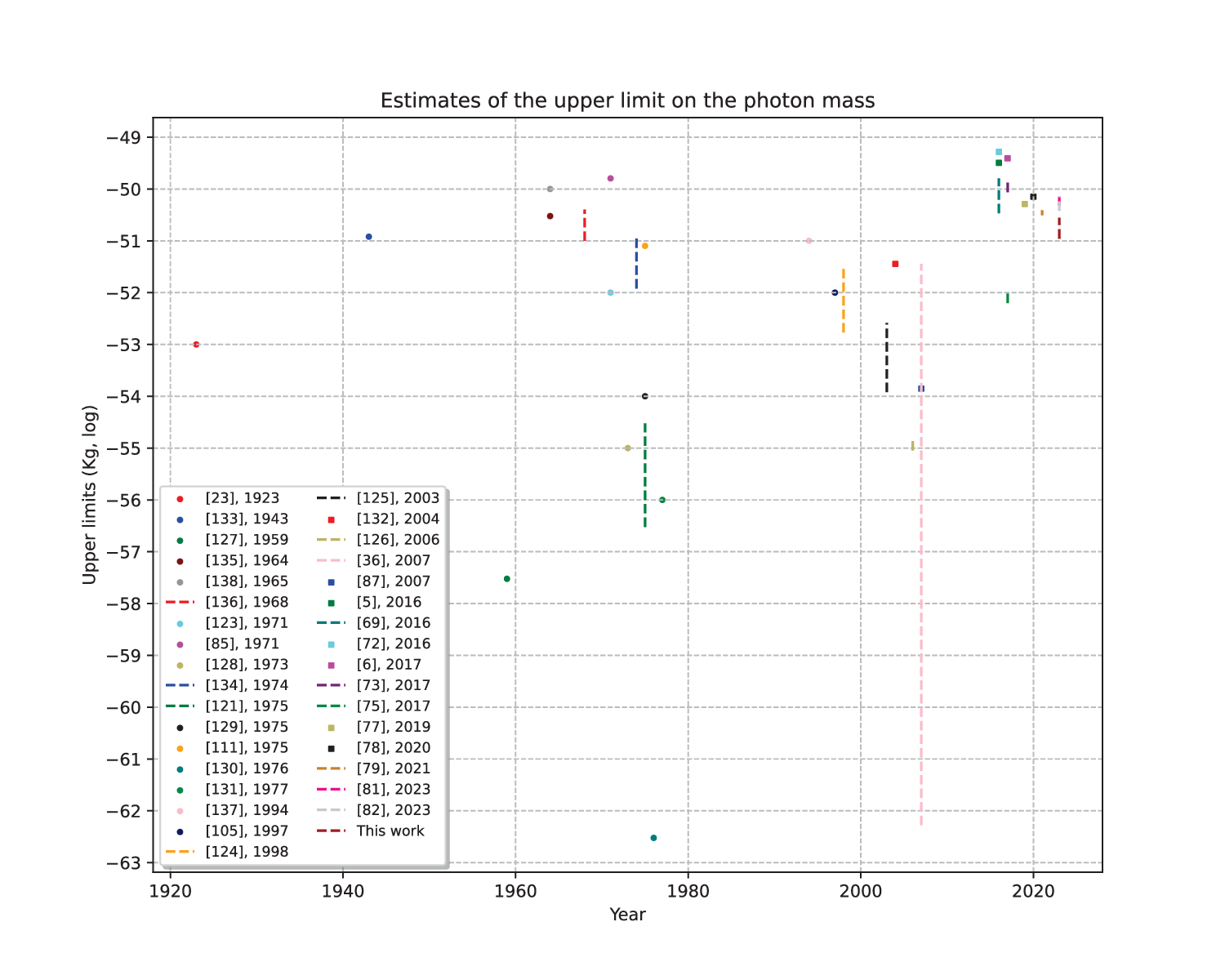}
        \caption{\footnotesize The upper limits listed in Tab. III plotted along the years.} 
    \label{fig8}
\end{figure}

\begin{table*}[!ht]
    \begin{tabular}{|p{8.7cm}|p{1.3cm}|p{7.65cm}
|}\hline
\multicolumn{3}{|c|}{Smallest (Heisenberg) measurable mass for any particle is $1.3\times 10^{-69}$ kg for $\Delta t$ = age of the Universe.}\\ \hline
Reference & Value \!(kg) & Method 
\\ \hline 
\citet{Barnes-Scargle-1975} & $3 \times 10^{-57} \newline 3 \times 10^{-56}$ & Observations of the Crab Nebula magnetohydro-dynamic waves 
\\ \hline \hline \hline
   \citet{Ryutov-2007} & $1.4 \times 10^{-54}$ & Model of the solar wind magnetohydro-dynamics (40 AU) \newline Official PDG limit.
\\ \hline \hline \hline
\citet{debernardis-etal-1984} & $2.9 \times 10^{-54}$ & Cosmic background dipole anisotropy 
\\ \hline 
\citet{debroglie-1923} & $10^{-53}$ & Dispersion  
\\ \hline
 \citet{yang-zhang-2017} & $6.3 \times 10^{-53} \newline 9.6 \times 10^{-53}$ & Observations of pulsar spin-down 
\\ \hline
\citet{Ryutov-1997} & $10^{-52}$ & Model of the solar wind magnetohydro-dynamics (1 AU). 
\\ \hline
\citet{Franken-Ampulski-1971} & $10^{-52}$ & Low frequency resonance circuits 
\\ \hline  
\citet{Lakes-1998, 
Luo-Tu-Hu-Luan-2003,
tu-shao-luo-luo-2006} & $9 \times 10^{-56} \newline 3.3 \times 10^{-52}$  & Torsion pendulum  
\\ \hline
\citet{Yamaguchi-1959,
Byrne-Burman-1973,
Byrne-Burman-1975,
Chibisov-1976,
Byrne-1977,
addvgr-2007} & $3 \times 10^{-63} \newline 3.6 \times 10^{-52}$
  & Model of the galactic potential 
\\ \hline
 \citet{Fullekrug-2004} & $3.6 \times 10^{-52}$ & Speed of lightning discharges in the troposphere  
\\ \hline
\citet{Davis-Goldhaber-Nieto-1975} & $ 8 \times 10^{-52}$ & Satellite data of Jupiter magnetic field 
\\ \hline
\citet{schrodinger-1943} &$ 1.2 \times 10^{-51}$
 & Earth and Sun magnetic fields 
\\ \hline
\citet{Hollweg-1974} & $1.1 \times 10^{-52} \newline 1.3 \times 10^{-51}$ & Model of Alfv\'en waves in the interplanetary medium 
\\ \hline
This work & $1.1 \times 10^{-51} \newline 2.8 \times 10^{-51}$ & AM law via MMS satellite data 
\\ \hline
\citet{Gintsburg-1964,
goldhaber-nieto-1968,
fikllalupe94} &$ 10^{-51} \newline 4 \times 10^{-51}$
 & Earth magnetic field with (satellite) observational data 
\\ \hline
\citet{retino-spallicci-vaivads-2016} & $3.4 \times 10^{-51} \newline 1.6 \times 10^{-50}$ & AM law via Cluster satellite data 
\\ \hline
\citet{wifahi71} & $1.6 \times 10^{-50}$ & Laboratory test on Coulomb's law 
\\ \hline  
\citet{Patel-1965} & $10^{-50}$ & Model of Alfv\'en waves in the Earth magnetic field. 
\\ \hline
 \citet{
boelmasasgsp2016,
wuetal2016b,
boelmasasgsp2017,
shao-zang-2017,
yang-zhang-2017, 
xing-etal-2019,
wei-wu-2020,
wang-miao-shao-2021,
Lin-Tang-Zou-2023,
Wang-Wei-Wu-LopezCorredoira-2023} & $3.1 \times 10^{-51} \newline 5.2 \times 10^{-50}$ & Fast Radio Bursts 
\\ \hline
  \end{tabular} 
    \\
{\caption {Summary of the main photon mass upper limits at and below $10^{-50}$ kg. For full reviews, see \cite{tulugi05,acciolynetoscatena2010a,scharffgoldhaber-nieto-2010,spqigiro11,wei-wu-2021,Workman-etal-2022}.
}}
\label{state-of-the-art}
    \end{table*}

\end{widetext}

\clearpage

\subsection*{Data availability}

Data will be made available on reasonable request.

\bibliographystyle{apsrev} 

\end{document}